# Asymmetric magnetism at the interfaces of MgO/FeCoB bilayers by exchanging the order of MgO and FeCoB

*Md. Shahid Jamal[1], Sadhana Singh[2], Arun Singh Dev[1], Neha Gupta[3], Pooja Gupta[3,4], Mukul Gupta[1], Olaf Leupold[5], Ilya Sergueev[5], V. R. Reddy[1], Dileep Kumar[1,*]*

1. UGC-DAE Consortium for Scientific Research, Indore, India.
2. IIT Hyderabad, Academic Block B-216, Kandi, Sangareddy, 502285, India
3. Raja Ramanna Centre for Advanced Technology, Indore 452013, India.
4. HBNI, Training School Complex, Anushakti Nagar, Mumbai 400094, India.
5. Deutsches Elektronen-Synchrotron DESY, Notkestraße 85, 22607 Hamburg, Germany.

*Email: dkumar@csr.res.in*

## Abstract:

Interfaces in FeCoB/MgO/FeCoB magnetic tunnel junction play a vital role in controlling their magnetic and transport properties for various applications in spintronics and magnetic recording media. In this work, interface structures of a few nm thick FeCoB layer in FeCoB/MgO and MgO/FeCoB bilayers are comprehensively studied using x-ray standing waves (XSW) generated by depositing bilayers between Pt waveguide structures. High interface selectivity of nuclear resonance scattering (NRS) under the XSW technique allowed to measure structure and magnetism at the two interfaces, namely FeCoB-on-MgO and MgO-on-FeCoB, yielding an interesting result that electron density and hyperfine fields are not symmetric at both interfaces. The formation of a high-density FeCoB layer at the MgO/FeCoB (FeCoB-on-MgO) interface with an increased hyperfine field (~34.65 T) is attributed to the increasing volume of FeCo at the interface due to boron diffusion from $^{57}$FeCoB to the MgO layer. Furthermore, it caused unusual angular-dependent magnetic properties in MgO/FeCoB bilayer, whereas FeCoB/MgO is magnetically isotropic. In contrast to the literature, where the unusual angular dependent in FeCoB based system is explained in terms of in-plane magnetic anisotropy, present findings attributed the same to the interlayer exchange coupling between bulk and interface layer within the FeCoB layer.

## Introduction:

Magnetic tunnel junctions (MTJs) based on FeCoB as a magnetic electrode and MgO as a tunneling barrier gained much attention because of their applications in random access memories and as a sensor in magnetic disc drives[1–3]. High TMR ratio obtained in the in-plane magnetized MTJs (i-MTJs) with CoFeB/MgO/CoFeB structures[4,5]. Despite the high TMR ratios, these devices suffer from low switching efficiency, thermal stability, and storage density. The discovery of inter-facial perpendicular magnetic anisotropy (PMA) in such a structure is an outstanding achievement, which possesses a high TMR ratio, relatively small spin-transfer torque (STT) switching current, and low magnetic damping[4–7].

In these systems, interfaces play a vital role in deciding the magnetic anisotropy and TMR ratio.[5,8] Recently, amorphous FeCoB and MgO interfaces have been responsible for unusual two-step hysteresis loops with higher coercivity near the magnetic hard axis[9]. The mechanism for such magnetism has been widely debated, with most commonly suggested mechanisms such as bond-oriented anisotropy (BOA)[10] and pair-ordering anisotropy (POA).[9,11] Recently, boron diffusion at the interface[9,12,13] has been found to be responsible for a stress-induced uniaxial magnetic anisotropy in the boron deficient FeCoB interface layer superimposed with the isotropic bulk FeCoB layer, and it creates unusual step hysteresis loop in the system. The role of interlayer exchange coupling between both layers is determined by separating magnetic contributions by fitting hysteresis loops with an adequate mathematical function[9]. In view of the facts, considerable temperature-dependent studies in the literature are also done to study interface magnetism and its role in optimizing the magnetic and transport properties of the multilayer system[5,14]. It has been established that with increasing temperature, nano-crystallization of FeCo starts at the interfaces due to the boron diffusion.[15,16] Unfortunately, most studies provided average information on all phenomena irrespective of the individual interfaces. In the recent past, there have been several studies where exchange in the order of ferromagnetic-non-magnetic layers is found responsible for the drastic change in the interface structure[17]. Namely, in the case of the Fe/Cr(Tb) multilayer, it is found that the root mean square (RMS) roughness of the two interfaces Fe-on-Tb(Cr) and Tb(Cr)-on-Fe are not equal.[17] Similar to the above Fe/Cr(Tb) multilayer, Wenbin Li et al. used x-ray reflectivity (XRR) and x-ray fluorescence (XRF) techniques under x-ray standing waves (XSW) produced by periodic multilayers to explain the asymmetry in the Ti/Ni/Ti trilayer.[18] They observed that the roughness of the Ti-on-Ni interface is 0.64 nm, and that of the Ni-on-Ti interface is 0.40 nm. Sagarika Nayak et al. studied the Fe/NiFe bilayer system by alternating the order of the magnetic layers by polarized neutron reflectivity (PNR).[19] They found that the interfacial magnetic moment increased by almost 18% when the NiFe layer was grown over the Fe layer. A considerable amount of studies have also been put into optimizing the CoFeB/MgO interface also to improve performance. However, despite the substantial technological importance of FeCoB and MgO-based systems, no study is available in the literature where special attention has been paid to understanding the interfaces (exchanging the order of ferromagnetic layer) to understand the unusual magnetization reversal in FeCoB and MgO-based systems.

Interface-resolved magnetic information of FeCoB-on-MgO or MgO-on-FeCoB interfaces is essentially required to get genuine information about these structures. The main difficulty is to perform depth-resolved magnetic properties or magnetism from the interface, independently from the bulk layer. Various lab-based techniques, such as superconducting quantum interface devices[20], magneto-optic Kerr effect (MOKE)[7,21], vibrating sample magnetometer[22], and nuclear magnetic resonance, are available[23] These techniques, unfortunately, give average information about the magnetic layer and do not have sufficient depth resolutions or fail to probe the true interfaces. Conversion electron Mossbauer spectroscopy (CEMS)[24], PNR[25], and x-ray magnetic circular dichroism[26] are considered more informative and powerful methods to analyze interface magnetism. However, each technique has its advantages and disadvantages[27], which primarily depend on the sample size, structure, and nature of magnetic studies. For example, in the case of the CEMS technique, depth selectivity can be achieved with a thin probe resonant layer embedded at a definite depth.[27–29] To study interface magnetism, a set of samples need to be prepared with a probe resonant layer ($^{57}$Fe) at different depth positions. However, full reproducibility is difficult in separate depositions. In the case of PNR, the relatively very long measurement time (several hours) due to reduced scatters per unit volume requires a relatively big sample size to have sufficient neutron scattering from the interface.[30] Nowadays, the third-generation synchrotron radiation source-based grazing incident nuclear resonance scattering (GINRS) technique is a powerful method[31–33]. High scattering yield and isotope selectivity make it possible to measure the magnetic depth profile of magnetic thin films and multilayers. The proximity magnetism of the surface and interfaces can be obtained precisely using isotope material within a reasonable time, even from a fraction of a monolayer.

In the present case, an attempt has been made to explore the unusual magnetic properties in FeCoB and MgO-based systems and their correlation with the interface in FeCoB/MgO and MgO/FeCoB bilayer structures. Samples are prepared by enriching the FeCoB layer with a $^{57}$Fe isotope to study interface magnetism using the isotope selective GINRS technique. To enhance scattering resonance yield from the $^{57}$Fe, GINRS measurements are performed under XSW conditions. For this purpose, $^{57}$FeCoB/MgO and MgO/$^{57}$FeCoB bilayers are deposited in the Pt waveguide structure, and interface selectivity was achieved due to the crossing of XSW antinodes with the interface at an appropriate x-rays incident angle. Prior to GINRS measurements, the correct position of the antinode (inside the cavity $^{57}$FeCoB/MgO and MgO/$^{57}$FeCoB) with respect to the $^{57}$FeCoB layer is monitored by performing nuclear and electronic Fe fluorescence under XSW. The combined study of MOKE and GINRS with x-ray fluorescence (XRF) has been adopted to develop a better understanding of the FeCoB/MgO interface (by exchanging the order of MgO and FeCoB layers), which could be helpful in the development of FeCoB/MgO based MTJs.

## 2. Experimental

A set of six bilayer samples, FeCoB(x)-on-MgO(6.5 nm) and MgO(6.5 nm)-on-FeCoB(x), where x = 4, 10 and 14 nm, were prepared. For GINRS measurements, the FeCoB layers were deposited by sputtering a $^{57}$Fe enriched (~99.999% pure) target with the composition $Fe_{43}Co_{40}B_{17}$. High-dense Pt layers of 30 nm and 2.5 nm are used respectively as a buffer and capping to protect the surface and to generate XSW between Pt waveguide structures. All samples were deposited using an Ar ion beam sputtering technique on Si(001) substrate at room temperature at a base pressure of about $1\times10^{-6}$ torr. The chamber was flushed with pure Ar a few times before deposition to reduce the oxygen and water vapor contamination. The targets were kept such that the incident beam makes 45° to the target normal. The substrate and target were kept parallel at 15 cm apart. Each target was pre-sputtered for 2 minutes to remove the contamination from the target surface. During deposition, the working pressure was maintained at about $1\times10^{-3}$ torr by keeping the Ar gas flow rate at 5 sccm. Respective deposition rates in all the bilayers for Pt, FeCoB, and MgO were 0.7, 0.4, and 0.3 Å/s. Both sets of samples are deposited under identical conditions, ensuring chemical homogeneity and consistency in composition. The final sample structures $Si_{sub}$/Pt(30)/**MgO(6.5)/ $^{57}$FeCoB(x)**/Pt(2.5) and $Si_{sub}$/Pt(30)/**$^{57}$FeCoB(x)/MgO(6.5)**/Pt(2.5) are designated as M-FCB$_x$ and FCB$_x$-M, respectively, further in the manuscript.

X-ray reflectivity (XRR) measurements have been performed on both samples by using a diffractometer (model D5000 of Siemens) with Cu-Kα radiation to get information about the layer thicknesses and rms interface roughnesses. MOKE measurements were performed to get magnetic information from the bilayer structures. The $^{57}$Fe nuclei in the samples were excited by a photon beam with an energy of 14.4 keV for the GINRS and XRF measurements, which were conducted at the nuclear resonance beamline P01 at PETRA III, Hamburg, Germany.[27,32] The synchrotron operated in the 40-bunch mode with a bunch separation of 192 ns. An avalanche photodiode detector was used for GINRS measurement, having a time resolution of ~ 1 ns, while a silicon drift detector was used for XRF measurement.

## 3. Results and discussion

Figure 1 (a & c) gives the x-ray reflectivity of both sets of multilayers (M-FCBx and FCBx-M; x = 4, 10 and 14 nm), as a function of momentum transfer vector $q = 4\pi\sin\theta/\lambda$, where $\theta$ being the angle of incidence. Reflectivity data has been fitted using Parratt's formalism[34] to get the thickness (d) and electron density ($\rho$) of the individual layers in the multilayer. The continuous curve gives the fitting of the reflectivity data yielding the multilayer structure with fitting parameters shown in table 1.

**Table1:** XRR fitting parameters such as; thickness and electronic density for both sets of samples M-FCB$_x$ and FCB$_x$-M using Parratt's formalism. The thickness of the buffer (Pt=30nm) and capping (Pt=2.5nm) layers are identical in all samples. Errors in layer thickness are ±0.1 nm.

| Layer | M-FCB$_x$ | | | | | |
|---|---|---|---|---|---|---|
| | d (nm) | | | $\rho$ (×10$^{-5}$ Å$^{-2}$) | | |
| | x=4 | x=10 | x=14 | x=4 | x=10 | x=14 |
| $^{57}$FeCoB$_{bulk}$ | 1.0 | 7.0 | 9.4 | 5.24 | 5.28 | 5.26 |
| $^{57}$FeCoB$_{interface}$ | 3.0 | 3.0 | 3.0 | 5.88 | 5.86 | 5.89 |
| MgO | 6.3 | 6.3 | 6.5 | 3.02 | 3.04 | 3.04 |
| FCB$_x$-M | | | | | | |
| MgO | 7.4 | 6.8 | 6.9 | 2.25 | 2.27 | 2.20 |
| $^{57}$FeCoB | 4.1 | 9.3 | 13.9 | 5.25 | 5.28 | 5.27 |

The corresponding scattering length density (SLD) profiles for all six samples, extracted from the fitting, are presented in Fig. 1(b & d). Interestingly, in the case of M-FCB$_x$ samples, the best data is obtained only after adding a high-density FeCoB layer (say $^{57}$FeCoB$_{interface}$ layer) of about ~ 3 nm at the $^{57}$FeCoB-on-MgO interface. The density of the $^{57}$FeCoB$_{interface}$ layer ($\rho \approx 5.88 \times 10^{-5}$ Å$^{-2}$) is ~11% higher than the bulk $^{57}$FeCoB$_{bulk}$ layer ($\rho \approx 5.28 \times 10^{-5}$ Å$^{-2}$). On the other hand, in the case of FCB$_x$-M samples, the best fit of the XRR patterns was obtained for the uniform density of the FeCoB layer.

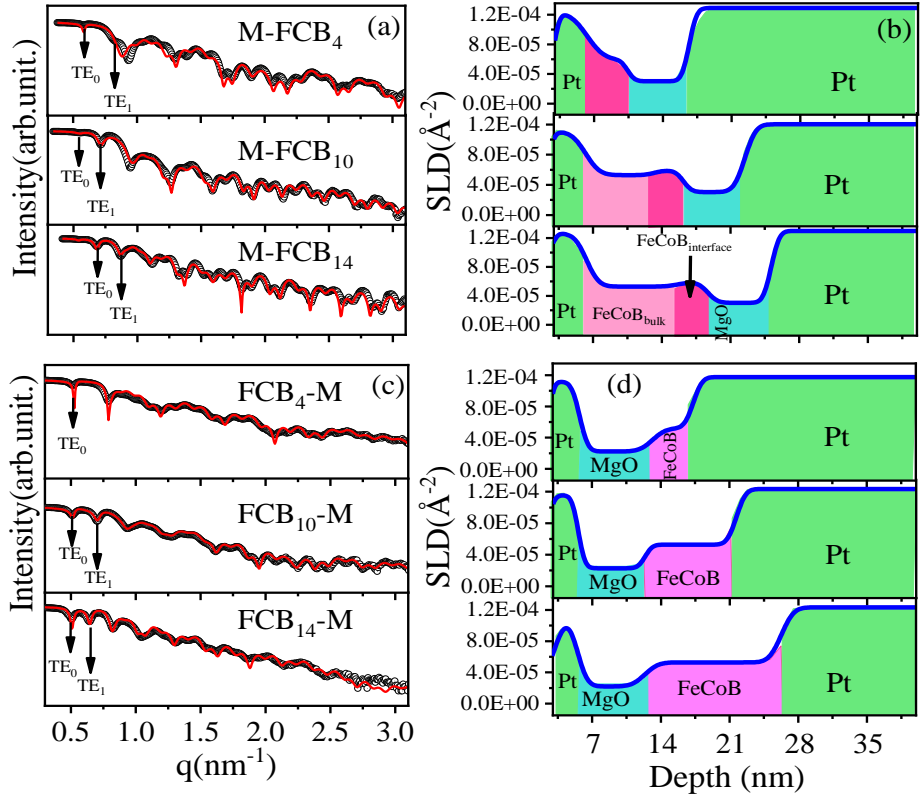

**Figure. 1**(a, c) X-ray reflectivity data (symbols) and corresponding best fit (solid red line) to the data for the samples M-FCB$_x$ and FCB$_x$-M by using Parratt's formalism, respectively. The corresponding SLD profile along with the depth of samples (b) M-FCB$_x$ (c) and FCB$_x$-M.

It is important to note that, in all sets of samples, sharp dips in XRR below the critical angle of Pt (q < 0.8 nm$^{-1}$) confirm the formation of standing wave modes in the cavity of the waveguide structure.[17,35,36] This aspect will be discussed later in the manuscript.

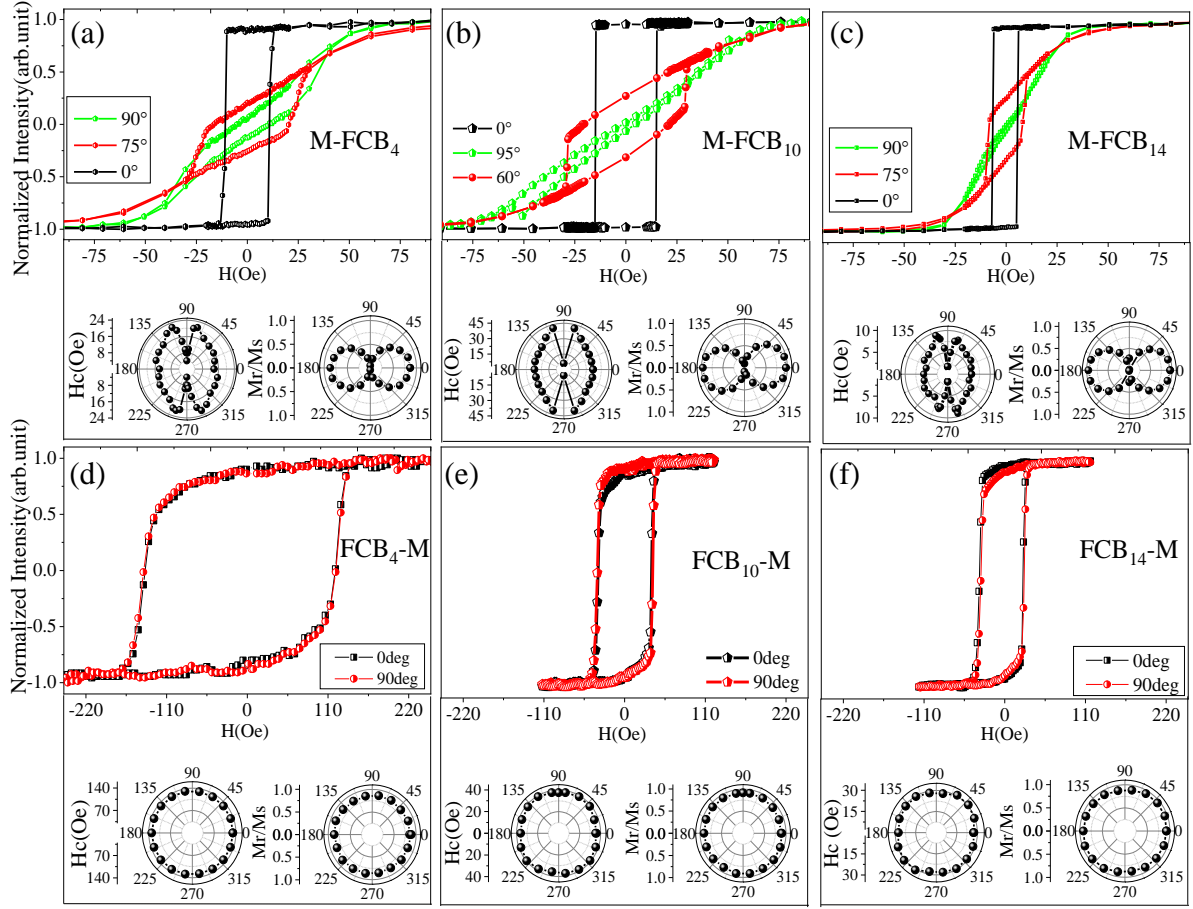

**Figure 2**. Representative MOKE loops with some azimuthal angles and the corresponding Hc and Mr/Ms plots as a function of azimuthal angle for samples (a-c) M-FCB$_x$ and (d-f) FCB$_x$-M.

Hysteresis loops of both sets of bilayer samples were collected as a function of azimuthal angle (ϕ) using MOKE. Figure 2 (a, b, & c) shows some representative hysteresis loops collected along ϕ = 0°, 75°, and 90°) for samples M-FCB$_{x=4, 10, 14 \text{ nm}}$. Along ϕ = 0° direction, the hysteresis loop is rectangular with a well-defined coercive field of ~15, 15, & 7 Oe for x = 4, 10, and 14 nm, respectively. With increasing ϕ, the rounding off of the hysteresis curve increases up to ϕ = 90° (hard axis), indicating the increasing contribution of the rotation of domain magnetization[9]. In polar plots with increasing ϕ, coercivity (Hc) and remanence (Mr/Ms) values are presented alongside of their respective MOKE hysteresis loops shown in Fig. 2(a-c).

The dumble shape of the Mr/Ms plots suggests the presence of UMA in all three samples of this set.[37] Interestingly, the Hc increases with increasing ϕ from 0° and reaches its maximum of about ϕ = 70-85°. On the other hand, Hc drops suddenly near zero around ϕ = 90°. In conventional magnetic anisotropy systems[37], which follow the Stoner-Wohlfarth model[9], Hc is always higher along the easy axis than in the other directions. Unexpected Hc behavior in the present case may be related to the high-density layer at the interface.

Contrary to this, as shown in figure 2 (d, e, & f), the hysteresis loops are almost independent of the azimuthal direction for all samples of set $FCB_x$-M. On the other hand, Hc is found sensitive to the increasing thickness of the $^{57}FeCoB$ layer. Hc and Mr/Ms vs $\phi$ are presented in polar plots along with their corresponding MOKE hysteresis loops in Fig. 2 (d-f). Hc reaches from 132 Oe to 28 Oe with increasing thickness of $^{57}FeCoB$ from 4 nm to 14 nm. Full range angular variation of Hc and Mr/Ms as a function of $\phi$ shows the absence of UMA.

To understand the origin of the unusual angular dependence of Hc in M-$FCB_x$ series, $H_c/H_a$ vs $\phi$ (Hc/Ha curve) are plotted in Fig. 3 (a-c). Here, $H_a$ is the anisotropic field constant, defined as the variation in the magnetic fields needed to reach saturation magnetization along the easy ($\theta = 0°$) and hard ($\theta = 90°$) magnetization direction.[38] It can be obtained with the help of easy and hard hysteresis loops using the following relation-

$$H_a = H_{sat}(\text{hard loop}) - H_{sat}(\text{easy loop}) \qquad 1$$

Where $H_{sat}$ is the saturation field, Ha is calculated 58, 66, and 35 Oe, respectively, for x = 4, 10, and 14 nm samples in series M-$FCB_x$ (see Fig. 2a to 2c) Hc/Ha curve compared with available models for magnetic anisotropy such as Stoner-Wohlfarth (SW) model[39], Kondorsky model[40], and two-phase model[41].

It is to be noted that for uniaxial systems in which the reversal mechanism is controlled by domain wall depinning, the *Hc* is in its simplest form described by the Kondorsky relation $H_c^{Kon} = Hc(0)/\cos\phi$.[40] It gives a monotonic increase in Hc as a function of the angle from 0° to 90°; hence, it can not match the observed Hc variation in the present case due to divergence at 90°.[42]

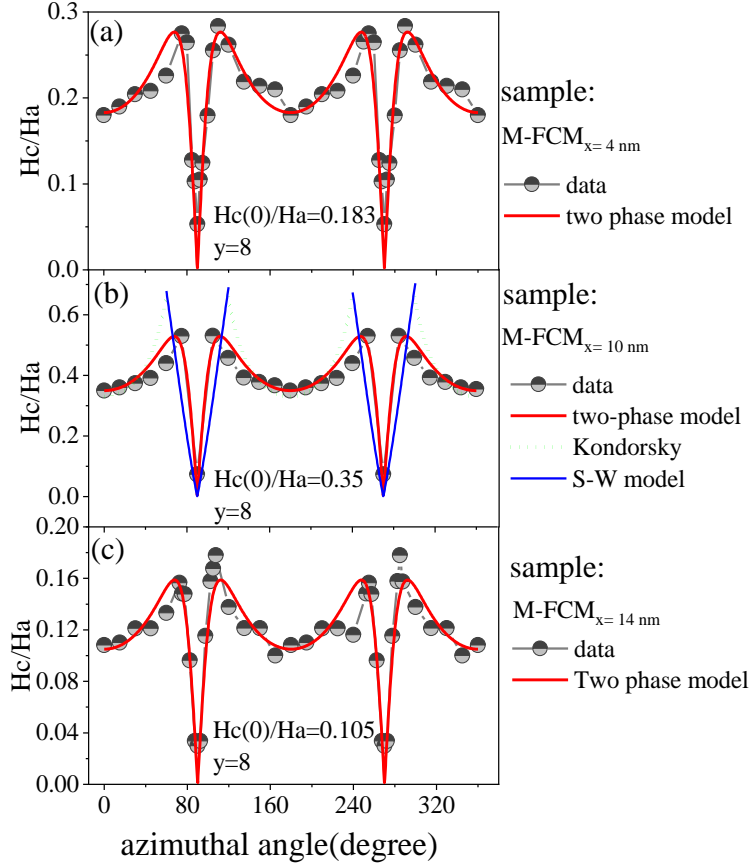

**Figure. 3.** (a-c) *Hc/Ha* variation as a function of azimuthal angle (ϕ) for series M-FCB$_x$ for x = 4, 10, and 14 nm, respectively and fitted with Eq. (2) for all samples. In (b), the fittings of the *Hc/Ha* vs ϕ, in the limited range of ϕ using Kondorsky (dashed green line) and S-W (solid blue line) models, are also shown, together with the fitting of the Hc/Ha vs ϕ in the entire range of ϕ (solid red line) using two phase model.

Stoner-Wohlfarth (S-W) model, which is based on coherent rotation, Hc is a monotonously decreasing function of ϕ; therefore, it will also not explain $H_c$ variation[39]. To demonstrate the same, Fig. 3(b) gives the fitting of the *Hc* curve for the "M-FCM$_{10}$" sample using the SW (blue line) and Kondorsky (dashed green line) models separately. But neither of the two models fit the data correctly.

In the case of the two-phase model, the magnetization reversal mechanism combines the domain wall nucleation (Kondorsky model) and domain rotation (S-W model)[41–44]. According to this model, the Hc variation as a function of ϕ is expressed as [44]

$$Hc(\phi) = \frac{Hc(0)\,(N_N+N_x)\cos\phi}{N_z \sin^2\phi + (N_N+N_x)\cos^2\phi} \qquad 2$$

For ϕ = 0°, applied magnetic field along the easy axis direction, the coercive field is equal to the nucleation field Hc(0). The $N_x$ and $N_z$ are the demagnetizing factors in the x-axis and z-axis direction, respectively (x-axis corresponding to the easy axis of magnetization). $N_N$ is the effective demagnetization factor defined in terms of anisotropy field $H_a$ and saturation magnetization Ms by the equation $N_N = H_a/M_S$. The parameter $y = (N_x+N_N)/N_z$ is a measure of anisotropy strength, and it can be obtained by fitting the above equation. For a large value of y, the above equation reduces to the original Kondorsky relation $H_c^{Kon} = H_c(0)/\cos\phi$[40]. The above equation predicts a maximum value of Hc close to the hard axis direction (70-85°) and zero for ϕ = 90°. Hence, the divergence in the hard axis direction of the Kondorsky relation is removed and replaced with a minimum. As shown in Fig. 3(a-c), the Hc/Ha variation is reasonably fitted (continuous red line) using a two-phase model for a whole angle ϕ with equation 2, indicating the validity of the two-phase model in explaining the magnetization reversal.

To further understand the unusual magnetization reversal precisely, interface resolved magnetism of two samples for x = 10 nm from both series (M-FCM$_{10}$ and FCM$_{10}$-M) is studied using GINRS under XSW conditions. Here, XSW is generated between the high-density Pt layers (cavity) due to the interference between the incident and reflected plane waves when x-rays are incident below a critical angle ($\theta_c$) of Pt. Rearrangement of the electric field intensity (EFI) in the cavity takes place such that several transverse electric modes are excited at fixed angle $\theta_m$.[35]

$$\theta_m = \frac{(m+1)\lambda}{2Wn} \qquad 3$$

where $\lambda$, $W$, n, and $m$ are the wavelength of x-rays, the thickness of the cavity, the refractive index, and the order of the mode, respectively. The schematic of XSW modes inside the cavity is shown in Fig. 4.

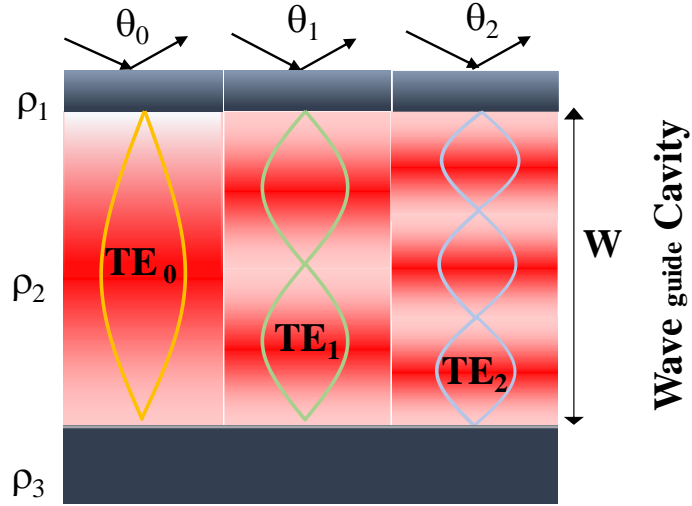

**Figure. 4.** Schematic representation of symmetric Pt waveguide structure. $W$ is the thickness of the cavity. $TE_0$, $TE_1$, and $TE_2$ are the transverse waveguide modes formed inside the cavity for $\rho_1 > \rho_2$ and $\rho_3 \geq \rho_1$, where $\rho_1$, $\rho_2$, and $\rho_3$ are the densities of top high-density Pt layer, cavity, and bottom Pt layer, respectively.

The formation of the EFI with an increasing incident angle inside the $^{57}$FeCoB/MgO cavity is simulated using Parratt's recursion algorithm[34] for samples $FCB_{10}$-M and $FCB_{10}$-M by taking an incident x-ray of energy 14.4 keV. As clear from the simulated contour plot in Fig. 5 (a & c), the resonance enhancement of the x-ray intensity inside the cavity occurs. The well-localized XSW antinodes corresponding to $TE_0$, $TE_1$, and $TE_2$ modes are visible in both the samples at a fixed value of $q_0 = 0.56$, $q_1 = 0.72$, & $q_2 = 0.96$ nm$^{-1}$ (for M-$^{57}$FCB$_{10}$) and $q_0 = 0.55$ & $q_1 = 0.73$ nm$^{-1}$ (for $^{57}$FCB$_{10}$-M), respectively. To establish how such a structure can improve the depth sensitivity from the $^{57}$FeCoB layer, XRF and nuclear resonance reflectivity (NRR) as a function of $q = 4\pi \sin \theta/\lambda$, is measured for both the samples and shown in Fig. 5(b & d) the contour plots. It is clear that the XRF and NRR spectrum consist of two peaks corresponding to the dips in XRR in both the samples at a fixed value of $q_0 = 0.56$ & $q_1 = 0.72$ nm$^{-1}$ for M-FCB$_{10}$ and $q_0 = 0.55$ & $q_1 = 0.73$ nm$^{-1}$ for $^{57}$FCB$_{10}$-M, respectively. The intensity of NRR from the thin layer is proportional to the fourth power of the standing wave amplitude compared to the square dependence of XRF.[45] Hence, the relative peak intensities for XRF and NRR are different.

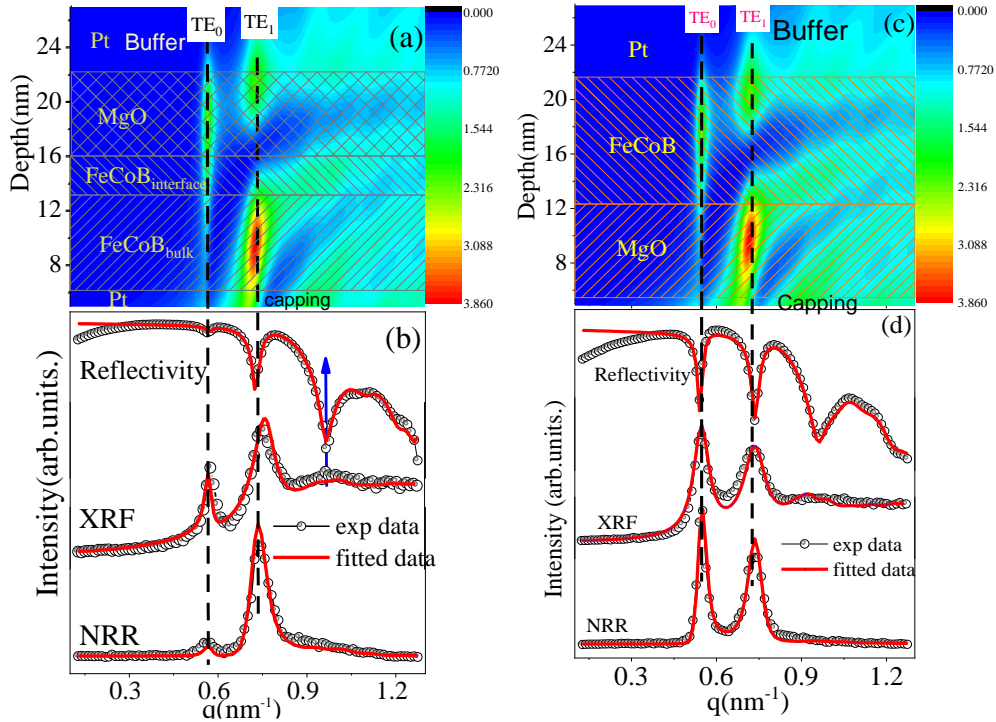

**Figure. 5.** The angular depth distribution of simulated intensity and fitted XRR, XRF and NRR pattern for sample M-FCB$_{10}$ (a&b) and FCB10-M (c&d). Position of the FeCoB and MgO layers are marked in the contour plot.

The origin of such peaks can be understood with the help of EFI in the trilayer, as shown in contour plots (Fig. 5 a & 5c). In the case of sample "M-FCB$_{10}$", for $q_0 \approx 0.56$ nm$^{-1}$, TE$_0$ antinode overlaps with mainly $^{57}$FeCoB-on-MgO interface layer (high-density layer), giving rise to the peak in the XRF and NRR at $q_0 \approx 0.56$ nm$^{-1}$. However, the increase in q, TE$_1$ antinode overlaps preferentially with the $^{57}$FeCoB$_{bulk}$ layer at $q_1 \approx 0.72$ nm$^{-1}$ and gives rise to the related peaks in XRF and NRR. For sample "FCB$_{10}$-M", at $q_0 \approx 0.55$, the TE$_0$ antinode fully overlaps with the whole $^{57}$FeCoB layer, whereas the TE$_1$ antinode overlaps partially with the $^{57}$FeCoB layer at $q_1 \approx 0.73$ nm$^{-1}$, giving rise to a relatively less intense peak in XRF and NRR curves compared to $q_0 \approx 0.55$ nm$^{-1}$. It is clear that a small variation in the depth and electron density of either part of the $^{57}$FeCoB layer would result in a significant variation in the relative intensities of fluorescence peaks corresponding to XSW modes.[27,32] Therefore, the experiment performed at these q values ($q_0$ and $q_1$) will enhance the contribution from the $^{57}$FeCoB layer (bulk and interface parts). Such enhancement in the signal becomes very important to investigate genuine properties when the probing layer is thin.

GINRS measurements were performed at q ≈ 0.56, 0.72, and 0.96 nm$^{-1}$ for sample M-FCB$_{10}$ and at q ≈ 0.55 and 0.73 nm$^{-1}$ for sample FCB$_{10}$-M. As we know, this technique is a time analogous to Mossbauer spectroscopy[31] and, therefore, sensitive to the hyperfine field and spin orientations of the resonant isotope ($^{57}$Fe) in probing layer $^{57}$FeCoB. The fitting parameters, such as the hyperfine field (Bhf), the width of hyperfine field distribution (ΔBhf), and the direction of the magnetic hyperfine field, are utilised for the fitting. The magnetic hyperfine field direction is defined in terms of two angles, namely, angle β with respect to the surface normal and the azimuthal angle γ with respect to the direction of polarization of x-rays (Fig. 6a). ΔBhf represents the distribution of hyperfine fields around the mean value Bhf. Oscillations in resonant count (fig. 6b and 6c) in GINRS spectra are due to the interference of electromagnetic waves emitted by different hyperfine components called quantum beats. The beating patterns in both the samples are fitted using REFTIM software[46] by considering the interface function as an "error function".

Figure 6(b) and 6 (d) represent the fitted GINRS spectra for samples M-FCB$_{10}$ and FCB$_{10}$-M. The best fit to GINRS data is obtained by taking bilayer structure as obtained from XRR, XRF and NRR measurements. For sample M-FCB$_{10}$, the four multiplets 29.85, 34.65, 32.43, & 22.41 T with (β, γ = 90°, 30°) having nuclear density concentrations of about 51, 31, 15, & 3 % in full $^{57}$FeCoB layer (including interface layer) are used and shown in fig. 6(c). It may be noted that GINRS simulated curve (dotted line) is also included in the same figure by considering the uniform $^{57}$FeCoB layer (no interface layer). Deviation from the best fit further confirms the existence of two magnetic layers within the FeCoB layer. The cause of different Bhfs contributions in the bulk and interface parts of the $^{57}$FeCoB layer is mainly due to the different magnetism caused by compositional differences near the interface. For sample FCB$_{10}$-M, the best fit to GINRS data (red lines) was obtained by taking five multiplets 29.88, 34.43, 32.45, 30.53, & 22.14 T with (β, γ = 90°, 10°) having nuclear density concentrations of about 48, 15, 16, 18, & 3 %, respectively, in whole $^{57}$FeCoB layer as shown in Fig. 6(e). Based on the best-fitted data for both the samples, the nuclear density distribution of Bhf along the depth of sample M-FCB$_{10}$ and FCB$_{10}$-M are presented in Fig. 6(c) and (e), respectively. The pie chart shows the nuclear density percentage contributions of Bhf for both samples in Fig. 7. For the M-FCB$_{10}$ sample, the total Bhf contributions for interface and bulk part, namely $^{57}$FeCoB$_{bulk}$ and $^{57}$FeCoB$_{interface}$, are separated after fitting and presented in table 2. For example, Bhf=34.65T is 31 % in the M-FeCoB$_{10}$ sample (see pie fig. 7), whereas 22

% of the same is at the interface layer. Similarly, it is found that Bhf of 29.85 T is present mainly with 29% in the $^{57}FeCoB_{bulk}$ layer, whereas the interface layer $^{57}FeCoB_{interface}$ contains only 22% of the part. Moreover, $^{57}FeCoB_{bulk}$ and the $^{57}FeCoB_{interface}$ layers contain 9% and 22% with Bhf of 34.25 T, respectively. The Bhf 32.43 T is almost uniformly distributed in both the layers ($^{57}FeCoB_{bulk}$:8%, $^{57}FeCoB_{interface}$:7%) while 22.41 T is present only in the $^{57}FeCoB_{interface}$ layer with 3%.

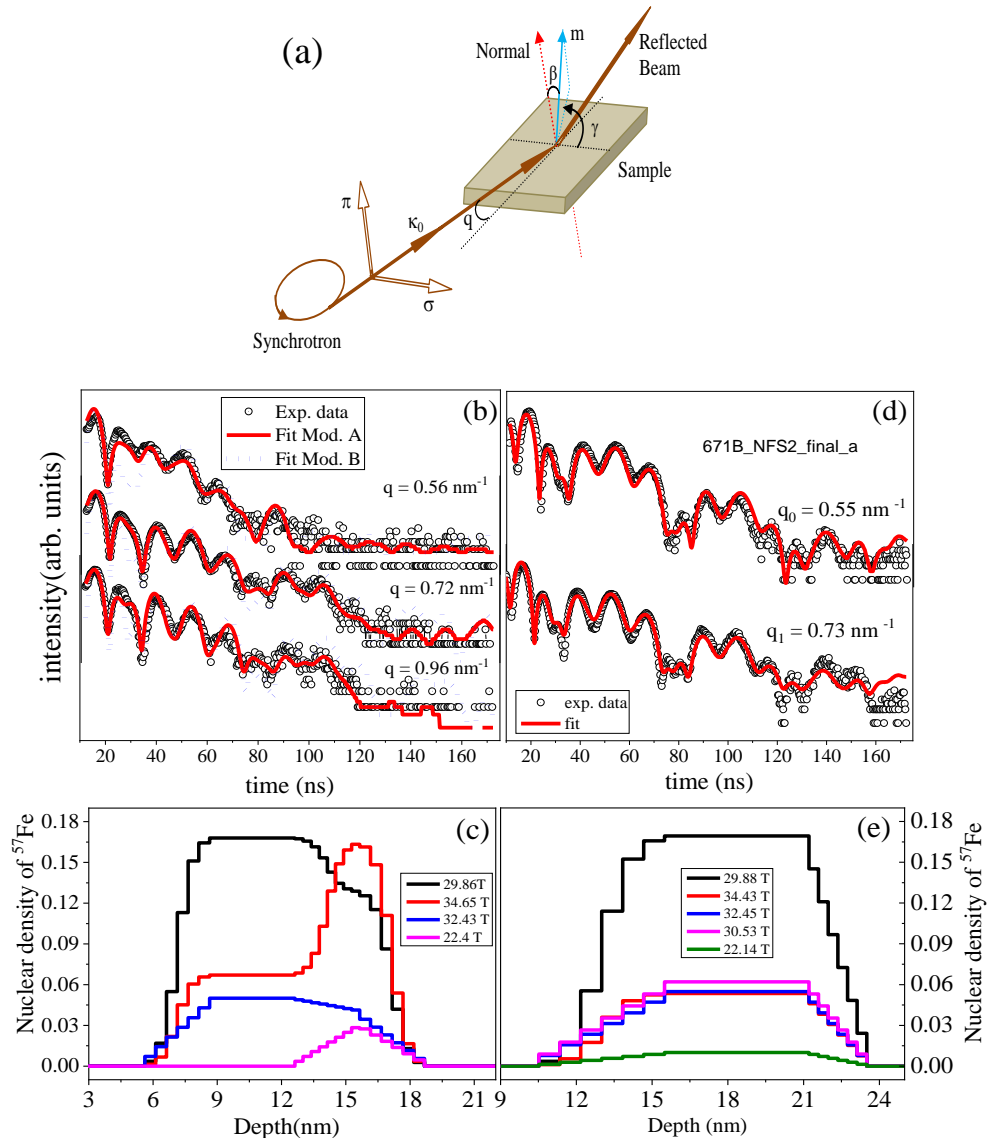

**Fig. 6(a)** Geometry used in the nuclear resonant scattering of synchrotron radiation from thin films and surfaces, defining the relative orientation (β, γ) of the incident wave vector $k_0$ to the direction of a unidirectional magnetization m of the sample. **(b) & (d)** GINRS data taken at different antinode positions of the standing wave, solid red lines are best fit to the data using REFTIM software. **(c) & (e)** represents

the multiplets density profile along with the depth for both the samples M-FCB$_{10}$ and FCB$_{10}$-M, respectively.

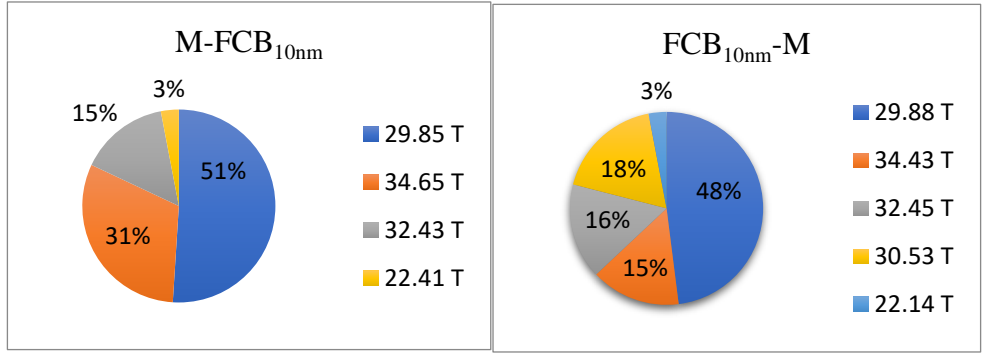

**Fig. 7.** Percentage contribution of Bhf in samples M-FCB$_{10}$ and FCB$_{10}$-M.

**Table 2:** Percentage contribution of different Bhfs in $^{57}$FeCoB$_{bulk}$ and $^{57}$FeCoB$_{interface}$ layer.

| Hyperfine fields | 29.85 T | 34.65 T | 32.43 T | 22.41 T |
|---|---|---|---|---|
| $^{57}$FeCoB$_{bulk}$ | 29% | 9% | 8% | - |
| $^{57}$FeCoB$_{interface}$ | 22% | 22% | 7% | 3% |

C. E. Johnson et al. investigated the Bhf in Fe$_{1-x}$Co$_x$ alloy with cobalt concentration, with the increasing amount of Co, the Bhf increases with pure iron and reaches the maximum value (36.5 T) with further addition of Co content and then decreases[47]. H. P. Klein et al. studied the variation of Bhf with Co content in (Fe$_{1-x}$Co$_x$)$_{80}$B$_{20}$, and they observed that Bhf varies between 25 to 28 T[48]. They also show that with a higher boron concentration, the Bhf decreases compared to the lower B content. In another study, two amorphous phases (with hyperfine fields 26.5 and 22 T) were observed in a Fe$_{40}$Co$_{40}$B$_{20}$ alloy sample of a thickness of 200 nm on a clean foil by DC magnetron sputtering with frequent interruption of plasma (100 times; for 30 s in every 2 nm layer) between the target and the substrate[49]. It is important to note that the different percentage contribution of Bhfs in $^{57}$FeCoB$_{bulk}$ and $^{57}$FeCoB$_{interface}$ layer in the M-FCB$_{10}$ sample is mainly due to compositional difference. A higher contribution of 29.85 T in the $^{57}$FeCoB$_{bulk}$ layer suggests the domination of the $^{57}$FeCoB phase. High ΔBhf in both the samples confirms the amorphous nature of $^{57}$FeCoB films[13]. In the case of the M-FCB$_{10}$ sample, the increased contribution of 34.43 T in the

$^{57}$FeCoB$_{interface}$ part and decreasing contribution of 29.85 T in the $^{57}$FeCoB$_{interface}$ region is mainly due to FeCo phase formation caused by boron diffusion during $^{57}$FeCoB deposition on MgO[9].

## Discussion

J. D. Burton et al. demonstrated using first-principle calculations that in the FeCoB/MgO/FeCoB MTJ, the presence of B causes the bonding of Fe-O or Co-O at the interface to be weakened by the existence of more advantageous Fe-B or Co-B bonding[50]. In another study by an x-ray photoemission spectroscopy of CoFeB/MgO bilayers were observed the formation of B, Fe, and Co oxides at the CoFeB/MgO interface due to oxidation of CoFeB during MgO deposition. In contrast to the literature, they observed that the vacuum annealing reduces the Co and Fe oxides formation but forms a composite MgB$_x$O$_y$ layer due to B diffusion into MgO[15]. Therefore, there is a contradiction regarding oxide formation at the FeCoB/MgO interface. In the present case, observation of a high-density layer at the interface negates the oxide layer formation; otherwise, a low-density layer with nonmagnetic contribution might have been observed in XRR and NRR data analysis.

It is important to note that the electronic density and Bhf of FeCo are higher than FeCoB layer[9,47,48]. Therefore, the only possible reason for the high-density interface layer ($^{57}$FeCoB$_{interface}$) and increasing contribution of higher Bhf at the interface in the M-FCB$_{10}$ sample is the diffusion of B from the $^{57}$FeCoB layer to the MgO layer near the interface. B migration into MgO can occur due to the higher electron affinity of Mg and O and its size which is smaller than Fe and Co. The MgO layer was deposited by the ion beam sputtering technique and has a finite probability of oxygen deficiencies[9,51,52]. The stress in the interface layer may stabilize the stress-induced UMA in the FeCoB$_{interface}$ layer at the interface[53,54]. Due to the lack of long-range structural order, the FeCoB layer generally does not possess UMA. The FeCoB$_{bulk}$ layer (isotropic) and FeCoB$_{interface}$ layer (anisotropic) are coupled magnetically and cause unusual angular-dependent hysteresis loops[9].

The observed angular-dependent shapes of the hysteresis loop are further verified and understood by depositing a separate trilayer structure (Si/Co(20 nm)/Alq3 (50 nm)/Co(20 nm)) with anisotropic and isotropic Co layer in the same trilayer sample. Before this, two separate samples, Si/Co (20 nm) and Si/Alq3(50 nm/Co(20 nm), were deposited in identical conditions.

It was found that Co exhibits magnetic anisotropy on the Si substrate (fig. 8a), whereas Co on Alq3 is isotropic. Anisotropy in Si/Co is mainly due to the long-range stress development during film growth.[53,54] On the other hand, these stresses are negligible or absent on the organic layer (Alq3) due to the mechanical softness of the Alq3 layer.

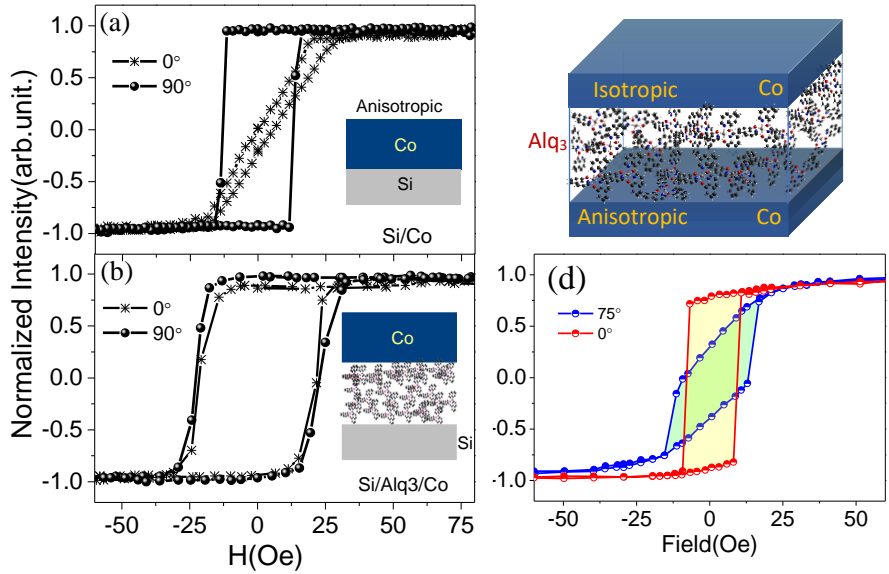

**Figure 8**. MOKE loops along 0° and 90° for the sample (a) Si/Co(20 nm), (b) Si/Alq$_3$(50)/Co(20 nm), (c) schematic of combined trilayer structure Co(20nm)/Alq3(50nm)/Co(20 nm), where top Co is isotropic, and bottom Co is anisotropic (d) MOKE loops of the trilayer along 0° and 75°.

Combining these two samples in a trilayer confirms the coexistence of anisotropic and isotropic layers in the same sample. Figure 8(d) gives MOKE loops of trilayer along its easy (0°) and near hard axis (75°) of magnetization. The hysteresis loop along 0° has Hc ~ 8 Oe, less than the Hc along ~75°. These loops match well with the M-FCB$_x$ bilayers in the present work. These findings further confirm and validate the model of the anisotropic interface layer in the M-FCB$_x$ bilayer. Here, the isotropic bulk and anisotropic interface layer couples and cause unusual shapes of hysteresis loops in azimuthal angle-dependent hysteresis loops.[9]

## Conclusion

The role of interface modification on the magnetization reversal of the $^{57}$FeCoB/MgO and MgO/$^{57}$FeCoB bilayer structures have been studied by performing interface resolved GINRS and XRF measurements under the x-ray standing waves (XSW) conditions. Highly sensitive interface resolved measurements allowed to measure structure and magnetism at the two interfaces, namely FeCoB-on-MgO and MgO-on-FeCo. The combined analysis revealed that electron density and hyperfine fields are not symmetric at both interfaces. $^{57}$FeCoB-on-MgO bilayer exhibits UMA, while MgO-on-$^{57}$FeCoB bilayer is found magnetically isotropic. The formation of a high-density FeCoB layer at the MgO/FeCoB (FeCoB-on-MgO) interface is attributed to the increasing volume of FeCo at the interface due to boron diffusion from $^{57}$FeCoB to the MgO layer. Furthermore, the formation of high dense layer at the interface caused unusual angular-dependent magnetic properties in MgO/FeCoB bilayer. In comparison to the literature, where the unusual angular dependent in FeCoB based system is explained in terms of in-plane magnetic anisotropy, present findings attributes the same to the interlayer exchange coupling between bulk and interface layer. The combined analysis revealed that the interchanging order of layers ($^{57}$FeCoB-on-MgO and MgO-on-$^{57}$FeCoB) plays a key role in tuning the magnetic anisotropy through interdiffusion at the interface.

## Acknowledgements:

A portion of this research was carried out at the light source PETRA-III of DESY, a member of the Helmholtz Association (HGF). Financial support by the Department of Science & Technology (Government of India) (Proposal No. I-20180885) provided within the framework of the India@DESY collaboration is gratefully acknowledged.